\documentclass[rnote]{aa} 
\usepackage{graphicx}
\usepackage{txfonts}
\usepackage{natbib}
\bibpunct{(}{)}{;}{a}{}{,} 

\begin{document}
%
  \title{On the backwards difference filter} 

   \author{R.~A. Garc\'\i a\inst{1}
   \and J. Ballot\inst{2}
	}
   \offprints{rafael.garcia@cea.fr}
   \institute {Laboratoire AIM, UMR 7158 CEA/DSM - CNRS - U. Paris Diderot, DAPNIA/SAp, 91191 Gif-sur-Yvette Cedex, France\\
\email{rafael.garcia@cea.fr}
   \and Max Planck Institut f\"ur Astrophysik, Karl-Schwarzschild-Str. 1, Postfach 1317, 85741 Garching, Germany\\
\email{jballot@mpa-garching.mpg.de}
        }

   \date{Received 6 March 2007 / Accepted 15 October 2007}

 
  \abstract
   {It is usual in helioseismology to remove unwanted instrumental low-frequency trends by applying high-pass filters to the time series. However, the choice of the filter is very important because it can keep the periodic signals throughout the spectrum. At the same time, these filters should not introduce any spurious effects on the remaining signal, which would modify the periodic signatures in the Fourier domain.}
   {One of the most used filters is the so-called backwards difference filter
   that can be applied when the time series are regularly sampled.  The
   problem of this filter is that the amplitudes and the phases of the periodic signals in the Fourier domain are modified and, therefore, their extracted amplitudes are biased when the frequency of the periodic signals decreases. The objective of this research note is to give a correction that could be applied to the power spectrum to over come this problem.}
   {We properly derive the transfer function of the backwards difference
   filter and we show how this can be applied to the power spectrum to correct
   the amplitudes. For the amplitude spectrum and the time series, we also
   derive a correction for the phase.}
   {The amplitudes of the periodic signals in the resultant power spectrum density are corrected from the cut-off frequency down to zero.  }
   {}


   \maketitle
%

\section{Introduction}
One of the most common problems to be faced when analyzing oscillatory components of an observed signal is the presence of low-frequency drifts which can be of instrumental or natural origin. These low-frequency trends introduce a background on the Fourier domain which reduces the signal-to-noise ratio (S/N) of searched periodic signals. To avoid this, high-pass filters are commonly used. The ideal filter is one that reduces the effect of the trend while preserving the periodic signals even at low frequencies. Moreover, it is important to keep all the parameters of periodic signals untouched, including their amplitudes.

In this research note we study a very easy-to-use filter: the backwards difference filter which is commonly used in helioseismology \citep[see for example][for a complete review on this research topic]{JCD2002} to reduce the low-frequency unwanted stochastic contribution of regularly sampled observations. We also deduce the transfer function of this filter. This transfer function allows us to correct the filtered power spectrum in order to retrieve the correct amplitudes of the periodic signals in the spectrum. Finally, we determine a correction for the resultant phase for when we are interested in the amplitude spectrum or in the time series.
 

\section{Backwards difference filter}

\subsection{Definition}
The backwards difference filter (hereafter called BDf) can be obtained by substituting the value of every data point of a regularly sampled time series by the difference of two consecutive points.
Let $\{f_n\}$ be a set of regularly sampled measurements of a function $f(t)$, such that $f_n=f(t_n)$ with $t_n=t_0+n\Delta t$, where $\Delta t$ is the sampling rate.
By definition, the filtered data set $\{\delta f_n\}$ is such that
\begin{equation}
\delta f_n = f_{n+1}-f_n
\end{equation}
which can be seen as the discretization of the function
\begin{equation}
\delta f(t)=f(t+\Delta t)-f(t) \label{eq:def}
\end{equation}
Thus, for a given $\Delta t$, the BDf can be defined as a linear operator, $\delta$, that can be applied to any function $f(t)$ using Eq.~\ref{eq:def}.
%

\subsection{Transfer function of the backwards difference filter}

Denoting $\widetilde{f}(\omega)={\cal F}[f(t)]$ the Fourier transform of the function $f(t)$, we can write that:
\begin{eqnarray}
\widetilde{\delta f}(\omega)
  &=& {\cal F}[f(t+\Delta t)-f(t)] \nonumber\\
  &=& [\exp(i\omega\Delta t)-1]\;\widetilde{f}(\omega) \nonumber\\
  &=& 2\sin\left(\frac{\omega\Delta t}{2}\right)
       \widetilde{f}(\omega)
       \exp\left(i\frac{\omega\Delta t+\pi}{2}\right)\label{eq:TFdf}
\end{eqnarray}
We deduce that the power spectrum of the original function $f(t)$ and the filtered one, $\delta f(t)$, are linked by:
\begin{equation}
|\widetilde{\delta f}(\omega)|^2=Q(\omega)\,|\widetilde{f}(\omega)|^2
\label{eq2}
\end{equation}
where $Q(\omega)$ is the transfer function of the BDf. It can be defined
from the previous expressions by:
\begin{equation}
Q(\omega)=\left[2\sin\left(\frac{\omega\Delta t}{2}\right)\right]^{2}
=\left[2\sin\left(\frac{\pi}{2}\frac{\omega}{\omega_c}\right)\right]^{2}
\label{eqq}
\end{equation}
where we have introduced the cut-off frequency $\omega_c=\pi/\Delta t$.
 \begin{figure}[htb*]
    \centering
    \includegraphics[width=\hsize]{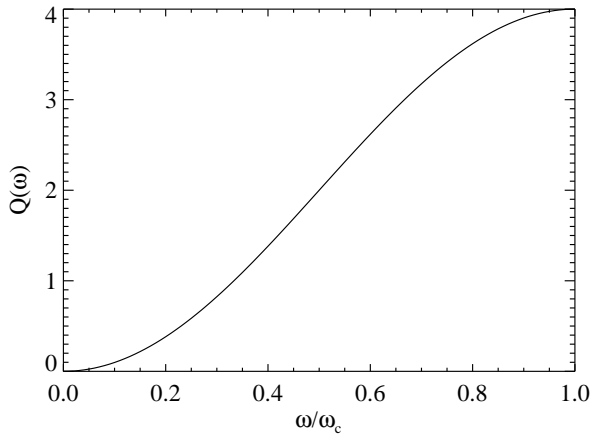}
    \caption{Transfer function $Q(\omega)$ (defined in Eq.~\ref{eqq}) of the backwards difference filter. The frequency $\omega$ has been normalized to the cut-off frequency $\omega_c=\pi/\Delta t$. \label{FigQ}}
\end{figure}

This transfer function shows the effect of the filter on the amplitude of the power spectrum (see Fig.~\ref{FigQ}). Indeed, at lower frequencies, the amplitudes tend towards zero. To recover the correct amplitudes in the filtered power spectrum, we need to divide it by $Q(\omega)$:
\begin{equation}
\mathrm{|\cal F}[f(t)]|^2=Q(\omega)^{-1}\, \mathrm{|{\cal F}}[\delta f(t)]|^2 
\label{eqqm}
\end{equation}

\section{Example}
\begin{figure}[htb*]
    \centering
    \includegraphics[width=\hsize]{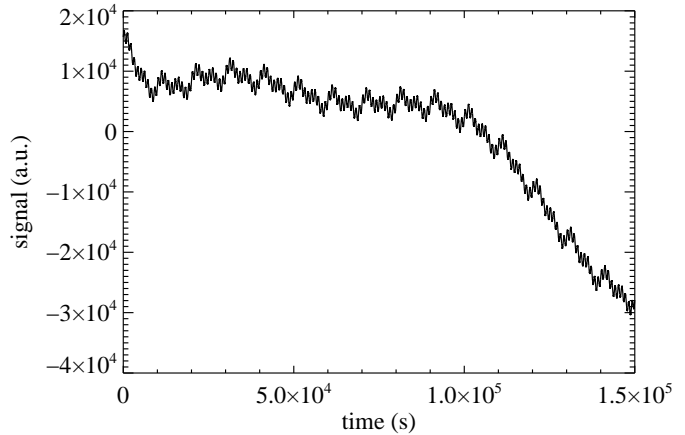}
    \caption{Simulated time series of the sum of three sine waves and a background that mimics a standard asteroseismic observed background.
    \label{FigTemp}}
\end{figure}

To illustrate the behavior of the BDf and to show the advantage of such a filter, we have simulated a simple observation containing low-frequency drifts and periodic signals (see Fig.~\ref{FigTemp}). The length of the simulated series is $T=150\,000\:$s with a constant sampling rate, $\Delta t =1\:$s. We have simulated three sine waves with periods of $1\,000$, $5\,000$ and $10\,000\:$s, corresponding to frequencies of $10^{-3}$,  $2\times 10^{-4}$ and $10^{-4}\:$Hz, in a way to cover a large range of frequencies. Their amplitudes are all equal to $1\,000$ arbitrary units. A low-frequency trend has been added to mimic the behavior of real observations. Thus, we have used a $7^\mathrm{th}$-order polynomial function fitted to the light curve of the Procyon data recovered by the MOST\footnote{A Canadian Space Agency mission, jointly operated by Dynacon Inc., the University of Toronto Institute for Aerospace Studies and the University of British Columbia, with the assistance of the University of Vienna} satellite. This data set can be found in the MOST public data archive\footnote{http://www.astro.ubc.ca/MOST/data/data.html} and it is described in \citet{MatKus04}.

\begin{figure}[htb*]
    \centering
    \includegraphics[width=\hsize]{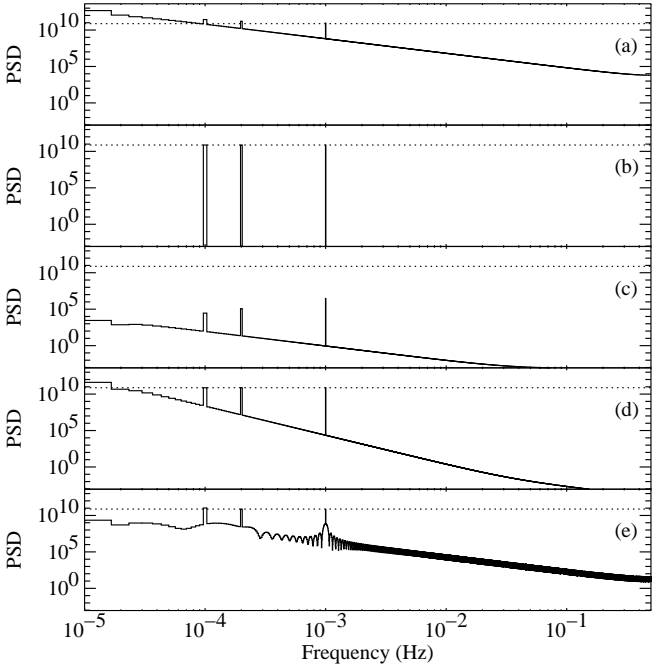}
    \caption{(a) Power spectrum density of the raw time series; (b) PSD of the three simulated sine waves without the low-frequency trend; (c) PSD after a direct (uncorrected) application of the BDf; (d) the same spectrum corrected by the $Q(\omega)$ factor; (e) PSD of time series filtered by removing a smoothing. The horizontal dotted lines correspond to the power of the simulated sine waves.\label{FigFiltro}}
\end{figure}

The power spectral density (PSD) -- calibrated as a single-sided spectrum, see \citet{1992nrfa.book.....P} -- of the simulated time series is plotted in Fig.~\ref{FigFiltro}a. The low-frequency trend introduces a background in the power spectrum reducing the S/N of the sine waves as the frequency decreases when it is compared to the PSD of the three sine waves computed without any trend (cf. Fig.~\ref{FigFiltro}b). The application of the BDf reduces the effect of the unwanted background (cf. Fig.~\ref{FigFiltro}c). Indeed the S/N ratio increase compared to the raw PSD. However, the amplitudes of the sine waves are modified by the transfer function of the filter. Thus, when we divide this PSD by the  $Q(\omega)$ factor, we retrieve the correct amplitudes (cf. Fig.~\ref{FigFiltro}d).

Another way sometimes used to filter slow trends is to remove a smoothing of time series.
To make a comparison, we have also plotted the PSD of our artificial data after applying such a filter. In this example, we have smoothed our time series by a boxcar function with a width of $T/10$ and subtracted it from the data.
Figure~\ref{FigFiltro}e clearly shows that, in this last case, the background of the PSD is dominated by wiggles which modify the background shape. Such wiggles can easily introduce artifacts.

In Table~\ref{table:1} we have summarized the S/N ratio of the three simulated sine waves. The S/N ratio has been defined as the ratio of the peak amplitude to the local background. It does not take into account any effects of stochastic noise in time series.
The increase in the S/N after applying the BDf is clearly shown.
For the data filtered by smoothing, the variations of S/N ratios are a direct consequence of the wiggles, and are very sensitive to the size of the boxcar which has been chosen.

\begin{table}[h]
\caption{S/N ratio in PSD of the three simulated sine waves at frequencies $\nu_1$= $10^{-4}$, $\nu_2$= $2\times 10^{-4}$  and $\nu_3$=$10^{-3}\:$Hz.}              
\label{table:1}      
\centering                                      
\begin{tabular}{l c c c}          
\hline\hline                        
Applied filter & \multicolumn{3}{c}{S/N} \\
& $\nu_{1}$ & $\nu_{2}$ & $\nu_{3}$ \\    
\hline                                   
    No (raw data) & 4 & 10 & 140 \\      
    BDf (corrected) & 314 & 5254   & $3 \times 10^6$ \\
    Smoothing & 165 & 268  & 98 \\
\hline                                             
\end{tabular}
\end{table}

\section{Correction to the phase}

We have seen that the BDf provides the right frequencies of the oscillatory signals and the right amplitudes after correcting them using the $Q(\omega)$ factor. However, the BDf also modifies the phase of the oscillations with a phase-shift depending on $\omega$. If we denote $\phi$ the real phase and $\psi$ the phase observed in $\delta f$, we can deduce a relationship between the two variables from equation (\ref{eq:TFdf}):
\begin{equation}
\phi = \psi-\frac{\omega \Delta t + \pi}{2}
     = \psi-\frac{\pi}{2}\left(1+\frac{\omega}{\omega_c}\right).
\end{equation}
The phase correction varies linearly from $-\pi/2$ to $-\pi$, over the whole
frequency range, from 0 to the cut-off frequency, $\omega_c$.

\begin{figure}[htb*]
    \centering
    \includegraphics[width=\hsize]{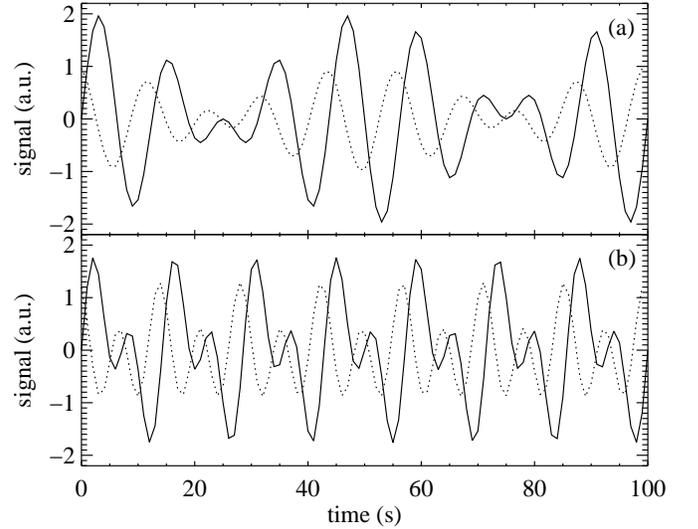}
    \caption{Examples of time series before (solid lines) and after
    (dotted lines) applying the BDf: (a) The signal contains two purely sinusoidal
    oscillations with close frequencies $\nu_0=0.07$~Hz and $\nu_1=0.08$~Hz;
    (b) The signal contains a purely sinusoidal oscillation with a frequency
    $\nu_0=0.07$~Hz and its first overtone ($\nu_1=0.14$~Hz). The sample rate
    is $\Delta t = 1$~s.\label{FigBeat}}
\end{figure}

The modification of the phase induces a modification of the shape of the
total signal. Figure~\ref{FigBeat} illustrates this phenomenon.
The first plot shows the beating of two oscillations with close
frequencies (0.07 and 0.08 Hz), that can be seen as a modulated carrier wave
of frequency 0.075~Hz. In the filtered series, the amplitude is, of course,
affected: it is multiplied by a factor $2\sin(0.075\pi)\approx0.47$. Furthermore
the wave has a phase lead of about $\pi/2$ (if we consider $\omega \ll \omega_c$).
The second plot shows what happens for a mode ($\nu_0=0.07$~Hz)
and its first overtone ($\nu_1=0.14$~Hz). We clearly see that the resulting
pattern is different. This strong modification appears because the phases have
been modified, even if the introduced
phase-shift for both oscillations is almost the same ($\psi_1\approx1.8$ and
$\psi_2\approx2.0$).

\section{Limitations in the application of the BDf}

As we have seen in the previous sections, the BDf can be a powerful detrender but it must not be overused. If there is no drift in the original signal, one has to avoid applying this filter. Indeed, there is always a small residual background in $\widetilde{\delta f}(\omega)$ which is amplified when we correct the amplitudes by dividing the original power spectrum by the $Q(\omega)$ factor. This effect is magnified at very low frequencies.

When the BDf is applied to any time series -- by construction -- we lose one
point of the signal (the last one) which is usually left at zero to maintain
the same number of points as in the initial series. Another problem could arise when data gaps are present in the original time series. When the BDf is applied to data containing N gaps, we lose N additional points as the consequence of the edges of those gaps. Therefore, if N is small compared to the total number of points, then the effect can be neglected. However if it is not the case, it has to be taken into account and if the number is too big, another kind of filter should be used.

\section{Conclusion}
In this research note we have shown the advantage of removing low-frequency trends by applying the backwards difference filter. The background in the PSD produced by this kind of trend is reduced and the S/N ratio of the periodic signals are improved. Moreover, the transfer function of this filter, $Q(\omega)$, has been deduced in order to correct -- by dividing the spectrum by this quantity -- the power of the periodic signals in the PSD. The resulting amplitudes inferred from the PSD are the original ones.

Filtering with a smoothing is often used, because it is a powerful detrending tool. However, it must be carefully applied and the choice of the boxcar size needs a preliminary study, according to frequencies of searched signals. The appearance of the wriggles must be taken into account, ideally by correcting the spectrum with the transfer function of this filter.

Even though this presentation does not deal with effects of stochastic noise, it emphasizes the advantages of the BDf and explains the manner to use properly this filter in a manner which is very easy to implement.

\begin{acknowledgements}
The authors want to thank the members of the asteroFLAG group%
\footnote{http://www.issi.unibe.ch/teams/Astflag/}
present at the first ISSI (International Space Science Institute) meeting for
triggering this discussion.
The authors want to thank the anonymous referee for useful comments that have improved the final version of this paper.
This work was partially supported by the European Helio- and Asteroseismology Network (HELAS\footnote{http://www.helas-eu.org/}), a major international collaboration funded by the European Commission's Sixth Framework Programme.

\end{acknowledgements}

\bibliographystyle{aa} 
\bibliography{/Users/rgarcia/Desktop/BIBLIO.bib}  

\end{document}